\documentclass[twocolumn,showpacs,amsmath,amssymb,superscriptaddress]{revtex4}

\usepackage{dcolumn}

\usepackage{graphicx}
\usepackage{dcolumn}
\usepackage{bm}

\begin{document}


\title{Spectroscopic calculations of the low-lying structure in exotic Os and W isotopes}

\author{K.~Nomura}
\affiliation{Department of physics, University of Tokyo, Hongo,
Bunkyo-ku, Tokyo, 113-0033, Japan} 

\author{T.~Otsuka}
\affiliation{Department of physics, University of Tokyo, Hongo,
Bunkyo-ku, Tokyo, 113-0033, Japan} 
\affiliation{Center for Nuclear Study, University of Tokyo, Hongo,
Bunkyo-ku Tokyo, 113-0033, Japan} 
\affiliation{National Superconducting Cyclotron Laboratory, 
Michigan State University, East Lansing, MI}

\author{R.~Rodr\'\i guez-Guzm\'an}
\affiliation{Instituto de Estructura de la Materia, CSIC, Serrano
123, E-28006 Madrid, Spain} 

\author{L.~M.~Robledo}
\affiliation{Departamento de F\'\i sica Te\'orica, Universidad
Aut\'onoma de Madrid, E-28049 Madrid, Spain}

\author{P.~Sarriguren}
\affiliation{Instituto de Estructura de la Materia, CSIC, Serrano
123, E-28006 Madrid, Spain} 

\author{P.~H.~Regan}
\affiliation{Department of Physics, University of Surrey, Guildford GU2
 7XH, United Kingdom}

\author{P.~D.~Stevenson}
\affiliation{Department of Physics, University of Surrey, Guildford GU2
 7XH, United Kingdom}

\author{Zs.~Podoly\'ak}
\affiliation{Department of Physics, University of Surrey, Guildford GU2
 7XH, United Kingdom}

\date{\today}

\begin{abstract}

Structural evolution in neutron-rich Os and W isotopes is investigated in
terms of the Interacting Boson Model (IBM) Hamiltonian determined by
(constrained) Hartree-Fock-Bogoliubov (HFB) calculations with
the  Gogny-D1S  Energy Density Functional (EDF). 
The interaction strengths of the IBM Hamiltonian are produced by 
mapping the potential energy surface (PES) of the Gogny-EDF with quadrupole
degrees of freedom onto the corresponding PES of the IBM system.  
We examine the prolate-to-oblate shape/phase transition which is predicted 
to take place in this region as a function of neutron number $N$ 
within the considered Os and W isotopic chains. 
The onset of this transition is found to be more rapid compared to 
the neighboring Pt isotopes. 
The calculations also allow prediction of spectroscopic variables (excited 
state energies and reduced transition probabilities) which are presented
for the neutron-rich $^{192,194,196}$W nuclei, for which there is only very
limited experimental data available to date. 

\end{abstract}


\maketitle



\section{Introduction\label{sec:introduction}}

Quadrupole collectivity has long been understood as one of the most
basic, yet prominent, aspects of nuclear structure \cite{BM,RS}. 
Nuclei are quantum many-body systems exhibiting collective 
properties associated with a distinct shape of the mean-field, which can be 
represented by a geometrical surface. Quadrupole collectivity can then 
be understood as a {\em quadrupole}-shaped
deformation of the nuclear surface whose magnitude depends  on the number
of {\em valence} nucleons, and has been shown to exhibit remarkable 
regularities in spectroscopic observables, such as the excitation energy of the 
$2_{1}^{+}$ state and the ratio $E_{4_{1}^{+}}/E_{2_{1}^{+}}$. Also evident 
are stunning shape/phase transitions at specific nucleon 
numbers where the collective nature of the quantal nuclear system can be well
described as a phase transition between (for example)  quadrupole vibrational and 
statically (quadrupole) deformed potentials \cite{RS,Casten_nature}.
 
The underlying multi-fermion dynamics of such nuclei however, is so
complex that its microscopic understanding 
still continues to be a theme of major interest in nuclear 
structure physics research. Mean-field studies, based on 
Skyrme \cite{Bender_review,Sk,VB} and Gogny \cite{Go} as well as 
relativistic \cite{Bender_review,Vretenar-1} 
Energy Density Functionals 
(EDFs) provide 
reasonable descriptions of various nuclear properties such as masses, charge
radii, mass density distributions, and surface deformations, over a wide
range of neutron and proton numbers \cite{RS,Bender_review}. Such  mean-field 
models, with their intrinsic spontaneous symmetry breaking mechanism, are highly 
relevant to understand the microscopy of the nuclear  quadrupole deformation and,
therefore have 
been used as a starting 
point for predictions relevant for future nuclear spectroscopic investigations
in exotic nuclei \cite{RS,Bender_review,CollSk,CollGo,CollRHB}.

Phenomenological studies using  the Interacting Boson Model
(IBM) \cite{AI} have enjoyed significant success in describing 
the low-lying quadrupole collective states of medium-mass and heavy
nuclei. 
The merit of the IBM lies in its simplicity, such that, based on 
group theory, the highly complicated multi-fermion dynamics of 
surface deformation can be simulated by simple, effective bosonic degrees 
of freedom, which correspond to (collective) pairs of valence 
nucleons \cite{OAI}. In addition to its success in reproducing 
a large amount of experimental 
data on low-lying collective nuclear states in heavy nuclei, 
the microscopic derivation of the IBM Hamiltonian has also been extensively
studied \cite{OAI,OCas,MO}. 
In particular, a novel way of deriving the interaction strengths of an
IBM Hamiltonian has been proposed recently \cite{nso}. 
This method is based on simulating the potential energy surface (PES) of a given EDF
by the corresponding IBM  PES. The IBM parameters are then derived 
as functions of the nucleon number using 
the Wavelet analysis method \cite{nsofull}. In this way, the universality of the
 nuclear EDF and the simplicity of
the IBM can be combined, thereby allowing the calculations to predict directly 
measurable spectroscopic observables such as excitation energies and 
electromagnetic transition rates between specific states. 
A number of spectroscopic calculations have been carried out 
using this method for the Ru, Pd, Ba, Xe, Sm isotopic chains, as well as 
theoretical predictions on $N>126$ Os-W nuclei, using the Skyrme EDF 
\cite{nsofull}.

The neutron-rich W, Os and Pt nuclei with $A\sim 190-200$ exhibit a 
very challenging structural evolution, which has already   been  extensively studied 
\cite{Shi06,Pod00,Jolie2003HfHg,Alk09,Reg08,Lane,Wheldon,Bond83}.
As originally pointed out in \cite{Pod00} the ratio
$E_{4_{1}^{+}}/E_{2_{1}^{+}}$ in $^{190}$W is anomalously 
small compared with the one in neighbouring isotopes. 
The most recent experimental data on the neutron-rich tungsten chain from 
$^{188,190,192}$W \cite{Shi06,Alk09,Lane} all suggest a change from a well
deformed, axially symmetric prolate shape for lighter tungsten isotopes, to a more 
gamma-soft system for $^{190}$W.
This transition from a prolate to very gamma-soft system for neutron
number $N=116$ (i.e., for $^{190}$W) is consistent with the recent
observation of the second 2$^+$ state in $^{190}$W which 
appears to lie lower than the yrast 4$^+$ in this nucleus \cite{Alk09}. 
The neutron-rich nature of the heavier W and Os nuclei make them
experimentally challenging to study. 
However, in recent years, there has been some progress in their
structural investigation following multi-nucleon transfer
\cite{Shi06,Lane,Wheldon} and isomer and/or beta-delayed gamma-ray
spectroscopy following projectile fragmentation reactions
\cite{Pod00,Alk09,Reg08}. 
The current experimental information is limited to the yrast sequence in
$^{190}$W \cite{Alk09,Lane} and the identification of the $2_{1}^{+}$
state in $^{192}$W \cite{Alk09}. 
It is interesting to note that the yrast 2$^+$ states in
the $N=116$ isotones $^{190}$W and $^{192}$Os  
have almost identical energies ($\sim$206 keV), as do the $N=118$
isotones $^{192}$W and $^{194}$Os ($\sim$218 keV).

On the theoretical side, mean-field calculations have been performed
which predict the shapes of these systems both with (e.g.,
\cite{stevenson}) and without (see, e.g., Refs.~\cite{RaynerPt,gradient-2}, and
references therein) the assumption of axial symmetry in the nuclear mean
field. 
The IBM has also been applied to fit the spectral properties of W isotopes 
in a phenomenological way \cite{DB_W}. 
More recently, spectroscopic calculations have been carried out \cite{nor}
to describe the structural evolution in Pt isotopes with the Gogny-D1S EDF
\cite{D1S}. In this paper, we review the current spectroscopy relevant
to the prolate-to-oblate shape/phase transition in neutron-rich Os and W
isotopes. 
We also report the predicted excitation spectra and
the transition probabilities on the neutron-rich Os and W nuclei.  
The spectroscopic calculations have been carried out in terms of the IBM
Hamiltonian derived by mapping  (constrained)  Hartree-Fock-Bogoliubov
(HFB) calculations, based on the Gogny-D1S EDF, using a
similar technique as in \cite{nor}.

\section{Theoretical procedures}
\label{sec:theory}

We begin with the calculation of the PES in terms of the (constrained)
HFB method using the Gogny-D1S EDF. 
The solution of the HFB equations, leading to the set of vacua
$|\Phi_{\rm HFB}\rangle$, is based on the equivalence of the HFB with a
minimization problem which is solved using the gradient method
\cite{gradient-2}. 
In agreement with the fitting protocol of the force, the kinetic energy
of the center of mass motion is subtracted from the Routhian  to be
minimized, in order to ensure that the center of mass is kept at rest. 
The exchange Coulomb energy is considered in the Slater approximation
and the contribution of the Coulomb interaction to the pairing field is
neglected. 
The HFB quasiparticle operators are expanded in a Harmonic Oscillator
(HO) basis having enough number of shells (i.e., $N_{shell}=13$ major
shells) to grant convergence for all values of the mass quadrupole
operators and for all the nuclei studied. 
The constraint is imposed on the average values of the mass quadrupole
operators 
$\hat{Q}_{20}=\frac{1}{2}\left(2z^{2}-x^{2}-y^{2}\right)$ and 
$\hat{Q}_{22}=\frac{\sqrt{3}}{2}\left(x^{2}-y^{2}\right)$ to the desired 
deformation values  
$Q_{20}=\langle\Phi_{\rm HFB}|\hat{Q}_{20}|\Phi_{\rm HFB}\rangle$
and 
$Q_{22}=\langle \Phi_{\rm HFB}|\hat{Q}_{22}|\Phi_{\rm HFB}\rangle$. 
In Refs.~\cite{RaynerPt,gradient-2}, the $Q-\gamma$ energy contour plots with 
$Q=\sqrt{Q_{20}^{2}+Q_{22}^{2}}$ and $\tan \gamma = Q_{22}/Q_{20}$
have been used to study the (mean-field) evolution of the ground state
shapes in Pt nuclei. 
Alternatively, one could also consider the  $\beta-\gamma$
representation in which the quadrupole deformation parameter $\beta$ is
written \cite{gradient-2} in terms of $Q$ as 
$\beta = \sqrt{\frac{4\pi}{5}} \frac{Q}{A \langle r^{2} \rangle}$, 
where $\langle r^{2} \rangle$ represents the mean squared radius evaluated
with the corresponding HFB state $|\Phi_{\rm HFB}\rangle$. 
The set of constrained HFB calculations provides the Gogny-D1S PES,
i.e., the total HFB energies $E_{\rm HFB}(\beta,\gamma)$. 

For the bosonic mapping we use the IBM-2, comprised of independent $L=0^{+},2^{+}$
proton ($s_{\pi}$, $d_{\pi}$) and neutron ($s_{\nu}$, $d_{\nu}$)
bosons. 
The number of proton (neutron) bosons, denoted by $n_{\pi}$ ($n_{\nu}$),
is equal to half of the number of valence protons (neutrons), assuming
the usual magic-number shell gaps at $Z=50$ and 82, and $N=82$ and 126. 
We adopt the standard IBM-2 Hamiltonian \cite{nso,nsofull,nor}. 
\begin{eqnarray} 
\hat H_{\rm IBM} &=& \epsilon(\hat n_{d \pi}+\hat n_{d \nu})+\kappa
 \hat Q_{\pi}\cdot \hat Q_{\nu}, 
\label{eq:bh}
\end{eqnarray}
where  $\hat n_{d \rho} = d_{\rho}^{\dagger}\cdot\tilde d_{\rho}$
and $\hat Q_{\rho} = [s_{\rho}^{\dagger}\tilde d_{\rho}+d_{\rho}^{\dagger}\tilde
s_{\rho}]^{(2)}+\chi_{\rho}[d_{\rho}^{\dagger}\tilde d_{\rho}]^{(2)}$
with $\rho=\pi,\nu$. 
The bosonic PES is represented by the expectation value of 
$\hat H_{\rm IBM}$ in the boson coherent state \cite{coherent}, given by 
\begin{eqnarray} 
|\Phi\rangle\propto\prod_{\rho=\pi,\nu}\Big[s_{\rho}^{\dagger}+\sum_{\mu=0,\pm 2}
\alpha_{\rho\mu}d_{\rho\mu}^{\dagger}\Big]^{n_{\rho}}|0\rangle   
\end{eqnarray}
where $|0\rangle$ stands for the boson vacuum (i.e., inert core) and the
coefficients $\alpha$'s are expressed as 
$\alpha_{\rho 0}=\beta_{\rho}\cos{\gamma_{\rho}}$, 
$\alpha_{\rho\pm 1}=0$ and 
$\alpha_{\rho\pm 2}=\frac{1}{\sqrt{2}}\beta_{\rho}\sin{\gamma_{\rho}}$. 
The intrinsic shape of the nucleus is then described in terms of the
(axially symmetric) deformation $\beta_{\rho}$ and the (triaxial) deformation
$\gamma_{\rho}$. The IBM PES reads \cite{nso,nsofull}
\begin{eqnarray}
& &E_{\rm IBM}(\beta_{\rm B},\gamma_{\rm B})=
\frac{\epsilon(n_{\pi}+n_{\nu})
 \beta_{\rm B}^2}{1+\beta_{\rm B}^2}+n_{\pi}n_{\nu}\kappa
 \frac{\beta_{\rm B}^2}{(1+\beta_{\rm B}^2)^2}\times \nonumber \\
& &\Big[4-2\sqrt{\frac{2}{7}}(\chi_{\pi}+\chi_{\nu})\beta_{\rm 
  B}\cos{3\gamma_{\rm B}}+\frac{2}{7}\chi_{\pi}\chi_{\nu}\beta_{\rm
  B}^2\Big], 
\label{eq:IBM-PES}
\end{eqnarray}
where $\beta_{\pi}=\beta_{\nu}\equiv\beta_{\rm B}$ and 
$\gamma_{\pi}=\gamma_{\nu}\equiv\gamma_{\rm B}$
is assumed for simplicity \cite{nso,nsofull}. We also assume the 
proportionality, i.e., $\beta_{\rm B}=C_{\beta}\beta$, where $C_{\beta}$ is a
numerical coefficient, and $\gamma_{\rm B}=\gamma$ \cite{nso,nsofull}. In
this context, the variables $\beta_{\rm B}$ 
and $\gamma_{\rm B}$ represent  the boson images of the (fermion) deformation 
parameters ($\beta$, $\gamma$). 
A point on the HFB PES, ($\beta$,$\gamma$), within an energy range relevant
for the considered low-lying quadrupole collective states, is mapped onto
the corresponding point on the IBM PES, ($\beta_{\rm B}$,$\gamma_{\rm B}$). 
The $\epsilon$, $\kappa$, $\chi_{\pi,\nu}$ and $C_{\beta}$ values are fixed
for a given nucleus by drawing the IBM PES so that the surface topology of
the corresponding HFB PES is reproduced. 
This is done unambiguously by means of the recently developed procedure 
\cite{nsofull} using the Wavelet transform \cite{wavelet}. 

Note that we compare the total energies $E_{\rm HFB}(\beta,\gamma)$
and $E_{\rm IBM}(\beta,\gamma)$. 
By reproducing the HFB PES as much as possible, effects of both
vibrational and rotational kinetic energies, similar to those introduced
when solving a five-dimensional (5D) collective Bohr Hamiltonian (see,
for example, Refs.~\cite{CollSk,CollGo,CollRHB}), should be included in
the boson systems.  
For large deformation, however, the rotational response, i.e.,
the response to cranking, differs significantly between nucleon and
boson systems, resulting in the deviation of the IBM rotational spectra
from fermionic ones. 
This deviation could be corrected by introducing an additional rotational
kinetic-like term, i.e., the so-called $L\cdot L$
term \cite{SU3} in the IBM Hamiltonian \cite{ibmmass}. 
This problem does not show up in the present work where only the
moderately deformed nuclei are concerned, and thus we neglect the
$L\cdot L$ term in the boson Hamiltonian of Eq.~(\ref{eq:bh}).  

We also note that to what extent the present mapping procedure mimics
the solution of a 5D Bohr Hamiltonian is an interesting open question,
which may be partly answered by looking at how reasonably our results
compare with the ones of the 5D Hamiltonian and the available
experimental data.

\begin{figure}[ctb!]
\begin{center}
\includegraphics[width=8.0cm]{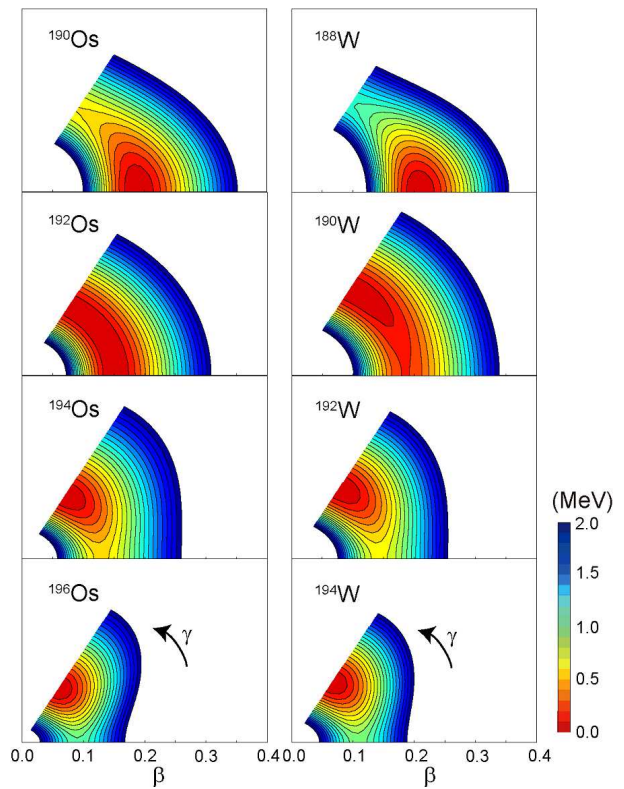}
\caption{(Color online) Mapped potential energy surfaces (PESs) for
 $^{190-196}$Os and $^{188-194}$W nuclei up to 2 MeV excitation from the
 energy minimum within the ranges 
$0^{\circ}\leqslant\gamma\leqslant 60^{\circ}$. 
The PESs are shown in terms of the fermionic
 deformation parameters $\beta$ ($=\beta_{\rm B}/C_{\beta}$) and
 $\gamma$ ($=\gamma_{\rm B}$).  }
\label{fig:pes}
\end{center}
\end{figure}

\section{Mapped IBM PES and the derived parameters}

Figure~\ref{fig:pes} shows the mapped IBM PESs for $^{190-196}$Os and
$^{188-194}$W nuclei up to 2 MeV excitation from the energy minimum. 
The corresponding HFB PESs have been reported in Fig.~3 of
Ref.~\cite{gradient-2}. 
The PESs for both Os and W nuclei show similar tendencies. 
There are quantitative differences between the Pt and 
Os-W isotopic chains, namely that the topology of the PES changes more
slowly in the former \cite{nor}, compared to the latter in Fig.~\ref{fig:pes}. 
An (almost) axially symmetric, 
oblate minima is observed in Pt nuclei with $N=114\sim 120$ and shallow
triaxiality for $N=110$ and 112 \cite{gradient-2,nor}. 
On the other hand, the Os and W isotopes are predicted to have the  
corresponding oblate minima only for $N=118$ and 120,
with a more rapid change to axially symmetric prolate deformation for
$N\leqslant 114$.
Indeed, shallow triaxiality (i.e., $\gamma$-softness) appears only
around $N=116$ for both Os and W nuclei \cite{gradient-2}. 
The corresponding mapped IBM PESs reproduce these trends of the
HFB PESs of \cite{gradient-2} well, whereas the location of the
minimum in the IBM PES differs from that of the HFB PES of
Ref.~\cite{gradient-2} in some nuclei as the present IBM PES of
Eq.~(\ref{eq:IBM-PES}) does not produce a triaxial minimum. 
The mapped PES for the $N=116$ isotone, $^{192}$Os is predicted to be very
flat along the $\gamma$-direction.
Similarly, the IBM PES for $^{190}$W is also very flat, with the global
energy minimum corresponding to a quadrupole deformation of 
$\beta\sim 0.15$ on the oblate side. 
This flatness is the consequence of the  $\chi_{\pi}$ and
$\chi_{\nu}$ parameter values, such that their sum is close to zero. 
Comparing Os and W isotopes with the same neutron number, the W nuclei
are generally steeper in both $\beta$ and $\gamma$ directions than the
corresponding Os isotone. 
A similar trend is also observed in the corresponding HFB 
PESs \cite{gradient-2}.

\begin{figure}[ctb!]
\begin{center}
\includegraphics[width=8.0cm]{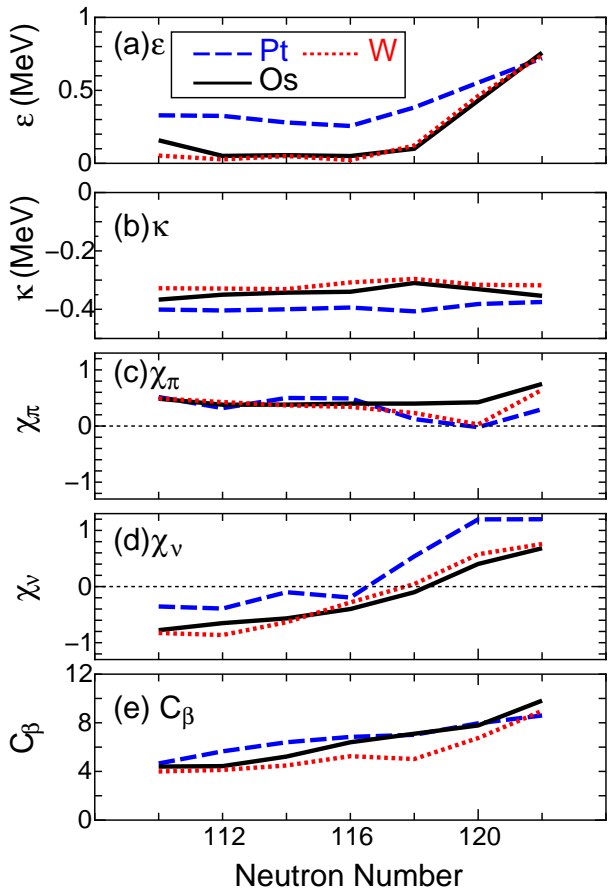}
\caption{(Color online) Derived IBM parameter values for the
considered Os and W nuclei, represented by solid and dotted
curves, respectively, as functions of $N$. 
Results for Pt isotopes taken from Ref.~\cite{nor} are also
 depicted for comparison.  }
\label{fig:para}
\end{center}
\end{figure}

Figure~\ref{fig:para} shows the evolution of the derived IBM parameters
for the considered Os and W nuclei as functions of the neutron number $N$. 
The parameter values for Pt nuclei, taken from Ref.~\cite{nor}, are also
shown for comparison. 
There are significant differences in
quantitative details of the derived IBM parameter values between Os-W and
Pt nuclei. 
In particular, the values of the parameter $\epsilon$ in
Fig.~\ref{fig:para}(a) for Os and W 
nuclei are rather small in the region away from the shell closure as
compared to Pt nuclei. 
In Fig.~\ref{fig:para}(b), the magnitude of the parameter $\kappa$ is
smaller than the analogous results for the Pt isotopes.  
The behavior of the parameters $\epsilon$ and $\kappa$ is reflective of
the HFB PESs for Os and W nuclei being somewhat steeper in the $\beta$
degree of freedom compared to the Pt isotopes, as discussed in
Ref.~\cite{gradient-2}. 
The $\chi_{\pi,\nu}$ parameters in Figs.~\ref{fig:para}(c) and
\ref{fig:para}(d) (as well as their sum) behave similarly to those of Pt
nuclei. 
For both $N$=110,112  the sum is almost zero in Pt isotopes while 
it is small for Os and W ones, but has a negative sign.
This indicates a weak prolate deformation in the latter 
as seen in Fig.~\ref{fig:pes}.  
In other words, the $\gamma$-soft structure is rather sustained in these Pt
isotopes, but it is not for the corresponding Os and W isotopes.  
As in Fig.~\ref{fig:para}(e), the scale parameter $C_\beta$ in the
present case behaves similarly as for the Pt nuclei with about the same
order of magnitude.

\section{Calculation of the energy spectra and $B$(E2) values}

Using the derived parameters, we calculate excitation spectra
and reduced E2 transition probabilities $B$(E2). 
The Hamiltonian of Eq.~(\ref{eq:bh}) is diagonalized by using the code
NPBOS \cite{npbos}.  

\begin{figure}[ctb!]
\begin{center}
\includegraphics[width=8.0cm]{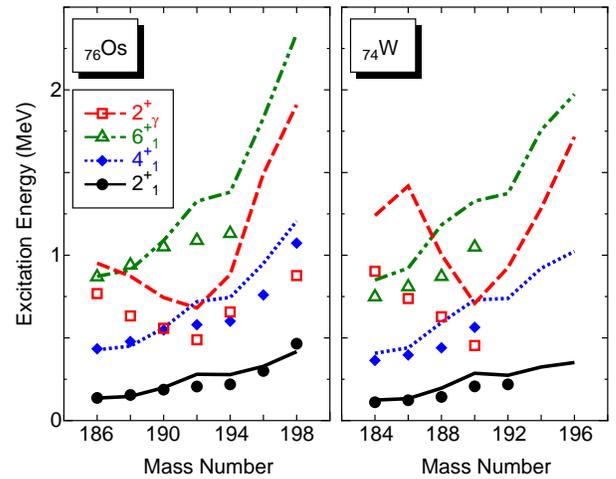}
\caption{(Color online) Low-lying $g.s.$ band and
 quasi-$\gamma$-bandhead  ($2^{+}_{\gamma}$) energies (curves) for
$^{186-198}$Os and  $^{184-196}$W nuclei. 
Experimental data (symbols) are taken from Refs.~\cite{Alk09,Pod09,data}. 
Gogny-D1S EDF is used. } 
\label{fig:osw}
\end{center}
\end{figure}

Figure~\ref{fig:osw} shows ground-state ($g.s.$) band and the
quasi-$\gamma$-bandhead $2^{+}_{2}$ (denoted by $2^{+}_{\gamma}$)
energies for Os and W isotopes.  
In general, the calculated results follow the experimental trends
reasonably well, particularly for $2^{+}_{1}$ energy. 
What is of interest in Fig.~\ref{fig:osw} is the behavior of the
$2^{+}_{\gamma}$ energy, exhibiting a kink 
for both $^{192}$Os and $^{190}$W. The experimental $2_{\gamma}^{+}$ energy 
in $^{192}$Os is lower than the $4^{+}_{1}$ one. 
This is an evidence that the $A=192$ nucleus is the most
$\gamma$-unstable one among other Os isotopes.  
The present calculation follows the trend for Os isotopes well, and
predicts a similar one for W isotopes exhibiting, however, more rapid 
change as a function of $N$. 
The location of the $2^{+}_{\gamma}$ state for $^{196}$Os ($^{192}$W) has
not yet been fixed experimentally but the present calculations suggest
that the $4^{+}_{1}$ state is lower than
the $2^{+}_{\gamma}$ one in both $^{196}$Os and $^{192}$W. 
The calculated $2^{+}_{\gamma}$ energy is generally higher than the
experimental one, whereas the qualitative feature of experimental level
is reproduced well.

Now we turn to the analysis of $B$(E2) systematics, relevant to the
considered low-lying states. 
The $B$(E2) value is given by 
\begin{eqnarray}
 B({\rm E2};J\rightarrow J^{\prime})=\frac{1}{2J+1}|\langle
  J^{\prime}||\hat T^{\rm (E2)}||J\rangle|^2, 
\end{eqnarray}
where $J$ and $J^{\prime}$ are the angular momenta for the initial and
final states, respectively. 
The E2 transition operator $\hat T^{{\rm (E2)}}$ is given 
by $\hat T^{\rm (E2)}=e_{\pi}\hat Q_{\pi}+e_{\nu}\hat Q_{\nu}$. 
Here $e_{\pi}$ and $e_{\nu}$ stand for the boson effective charges. 
These effective charges should be in principle determined 
not at the mean-field level, but rather by some treatment taking 
into account effects beyond the mean field, such as core polarization.  
This is, however, beyond the scope of the current framework and may need
to be investigated in the future. 
In what follows, we assume $e_{\pi}=e_{\nu}$, for
simplicity, and discuss ratios of $B$(E2)s rather
than their absolute values and the quadrupole moments for
the corresponding excited states. 
Note that the $B$(E2) ratio at each dynamical
symmetry limit, shown below, means the one with infinite boson number
\cite{AI}.

\begin{figure}[ctb!]
\begin{center}
\includegraphics[width=8.0cm]{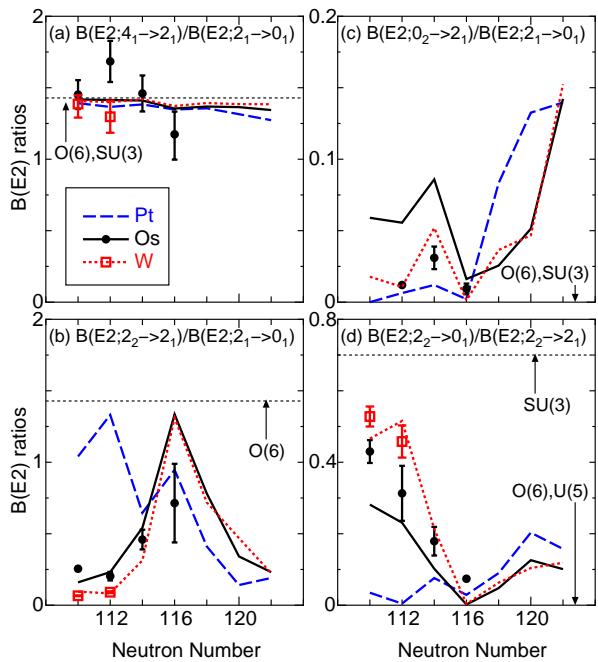}
\caption{(Color online) Theoretical (curves) and experimental 
(symbols with error bar) \cite{BE2} $B$(E2) ratios for Os and W
isotopes as functions of $N$. Theoretical results for Pt isotopes
taken from Ref.~\cite{nor} are also depicted as dashed curves, 
for comparison. Gogny-D1S EDF is used. }
\label{fig:BE2}
\end{center}
\end{figure}

\begin{figure*}[ctb!]
\begin{center}
\includegraphics[width=17.0cm]{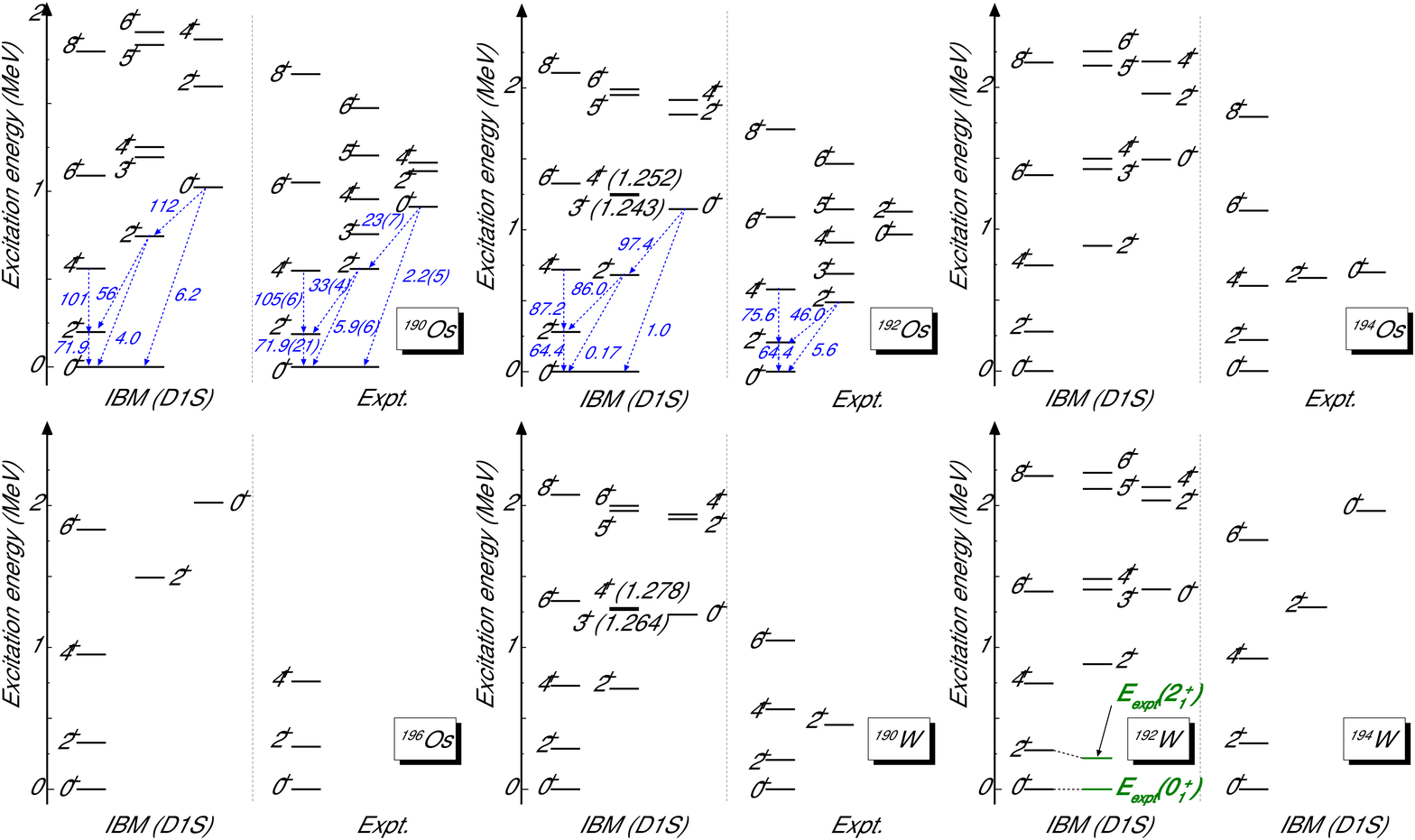}
\caption{(Color online) Level schemes for $^{190,192,194,196}$Os
 and $^{190,192,194}$W nuclei. The theoretical $B$(E2) values (in
 Weisskopf units) for $^{190,192}$Os are normalized to the experimental
 \cite{BE2} $B$(E2;$2^{+}_{1}\rightarrow 0^{+}_{1}$) value. 
Calculated $3^{+}_{1}$ and $4^{+}_{2}$ energies for $^{192}$Os and
 $^{190}$W are shown (in MeV units) in the parentheses because otherwise 
 these energies look
 identical. Note that, for $^{192}$W, experimental data are shown in
 the same panel as theoretical ones. Gogny-D1S EDF is used. } 
\label{fig:levelscheme}
\end{center}
\end{figure*}

From Fig.~\ref{fig:BE2}(a), we observe that the ratio 
$R_{1}\equiv B({\rm E2};4_{1}^{+}\rightarrow 2_{1}^{+})/B({\rm E2};
2_{1}^{+}\rightarrow 0_{1}^{+})$ 
does not change much, being close to its O(6) limit of IBM $10/7$ (which
is also the SU(3) limit of $R_{1}$). 
This trend persists for $N\geqslant 118$ where there is currently no
available data. 
The ratio 
$R_{2}\equiv B({\rm E2};2^{+}_{2}\rightarrow 2^{+}_{1})/B({E2};2^{+}_{1}
\rightarrow 0^{+}_{1})$, 
shown in Fig.~\ref{fig:BE2}(b), is of particular interest as one 
can observe a significant difference in its value for the Pt and Os-W
isotopes. 
The magnitude of the $R_{2}$ ratio is 
arguably the most appropriate and sensitive fingerprint for $\gamma$
softness \cite{nor}. 
The $R_{2}$ values for both Pt and Os-W are relatively large and 
close to the O(6) limit ($=10/7$) for $N=114\sim 118$, where the nuclei
show notable $\gamma$ instability. 
For Pt nuclei, this trend persists even for $N\leqslant 112$, while 
smaller values are suggested for  Os and W nuclei. 
These differences between the Pt and Os-W chains reflect the difference in
the topology of the PES. 
The results for Os nuclei follow the experimental trend, which increases 
for $N=110-116$.
The present calculation for Os nuclei suggests the decrease of the $R_{2}$
value for $N\geqslant 118$, which corresponds to a suppression of $\gamma$
softness. 
The ratio 
$R_{3}\equiv B({\rm E2};0_{2}^{+}\rightarrow 2_{1}^{+})/B({\rm E2};2_{1}^{+}
\rightarrow 0_{1}^{+})$ 
in Fig.~\ref{fig:BE2}(c) generally has a predicted value which is rather
small, being close to zero (corresponding to the O(6) and SU(3)
limits), as compared to $R_{1}$ and $R_{2}$ values. 
Note that the scale of the vertical axis in Fig.~\ref{fig:BE2}(c) is
different from those of Figs.~\ref{fig:BE2}(a) and \ref{fig:BE2}(b). 
No rapid change with $N$ is seen for $R_{3}$ as in $R_{2}$. 
Nevertheless, we should note the quantitative differences between the Pt
and the Os-W nuclei. 
The branching ratio 
$R_{4}\equiv B({\rm E2};2_{2}^{+}\rightarrow 0_{1}^{+})/B({\rm E2};2_{2}^{+}
\rightarrow 2_{1}^{+})$ 
in Fig.~\ref{fig:BE2}(d) for Os follows the experimental trend for $N=110\sim 116$. 
The decrease of $R_{4}$ value from $N=110$, close to $7/10$ (SU(3) 
limit), toward $N=116$, close to zero (O(6) and U(5) limit), reflects 
the corresponding structural evolution. 
The $R_{4}$ value for the Pt chain is close to zero, while for the Os-W
chains, there is a significant change at $N=116$. 
For the W nuclei, the ratio $R_{4}$ increases more rapidly than for the Os chain from
$N=116$ to 112. 
Earlier phenomenological studies suggested a similar increase \cite{DB_W}.

Finally, we present in Fig.~\ref{fig:levelscheme} the 
level schemes corresponding to the 
neutron-rich nuclei  $^{190,192,194,196}$Os and $^{190,192,194}$W
taken as representative samples. For $^{190,192}$Os, for which there
are significant experimental data, 
not only the $g.s.$ band but also both the quasi-$\gamma$-bandhead
$2^{+}_{\gamma}$ and the quasi-$\beta$-bandhead $0^{+}_{2}$ 
(denoted by $0^{+}_{\beta}$) energies are reproduced quite well by the
current calculations, although the detailed 'in-band' energy staggering looks
different between the calculated and the experimental levels. 
The calculated $B$(E2) values for $^{190,192}$Os have been normalized to 
the experimental \cite{BE2} $B$(E2;$2^{+}_{1}\rightarrow 0^{+}_{1}$) value. 
Some algebraic feature is also apparent in the calculated results.  
The $\Delta\tau=\pm 1$ rule for the E2 decay pattern at the O(6) limit 
\cite{AI}, (i.e., the dominance of $2^{+}_{2}\rightarrow 2^{+}_{1}$ 
($0^{+}_{2}\rightarrow 2^{+}_{2}$)
over $2^{+}_{2}\rightarrow 0^{+}_{1}$ 
($0^{+}_{2}\rightarrow 2^{+}_{1}$)) in the present calculation also compares
well with the experimental decay pattern.

The experimental value of the $2^{+}_{1}$ energy for $^{192}$Os is 
very close to that of its isotone  $^{190}$W (i.e., $E \approx$207 keV).  
Also, the excitations energies of the $2^{+}_{1}$ levels in these
isotones are also quite similar to each other. 
The present calculations reproduce this overall trend well. 
In fact, the calculated $E(2^{+}_{1})=0.280$ (0.278) MeV
and 0.286 (0.274) MeV for $^{192}$Os ($^{194}$Os) and $^{190}$W
($^{192}$W) nuclei, respectively. 
For $^{192}$Os and $^{190}$W nuclei, the calculated $g.s.$ band energies
are rather stretched, and the $2^{+}_{\gamma}$ energies are in good
agreement with the respective experimental data. 
In the calculated quasi-$\gamma$ band of $^{190}$Os and
$^{192}$Os nuclei, one observes a staggering as 
$2^{+}_{\gamma}$ ($3^{+}_{\gamma}$ $4^{+}_{\gamma}$) ($5^{+}_{\gamma}$
$6^{+}_{\gamma}$), ... etc.. 
By contrast, the experimental energy spacing shows a more regular pattern. 
This deviation may be related to the topology of the mapped IBM PES
in Fig.~\ref{fig:pes}, which is flat in $\gamma$ direction,
while the corresponding Gogny-D1S PES exhibits shallow triaxial minimum
\cite{gradient-2}. 
In the future, some additional interaction term, such as a
so-called cubic term \cite{cubic}, may need to be introduced in the boson 
Hamiltonian to correct the deviation for detailed structure of 
quasi-$\gamma$ band.

For $^{194,196}$Os nuclei, the predicted $2^{+}_{1}$ and $4^{+}_{1}$ 
energies reproduce the experimental ones. 
The quasi-$\beta$-bandhead energy for $^{194}$Os in the present calculation 
is notably larger than the experimental value, which is a consequence of 
the peculiar topology of the Gogny-HFB PES,
which exhibit a pronounced oblate minimum with a relatively small 
deformation. 
This results in the larger value of the parameter $\kappa$ than the one
in the IBM phenomenology \cite{DB_W} which would give good agreement for
the excited $0^{+}$ energies. 
The positions of the $2^{+}_{\gamma}$ and the $0^{+}_{\beta}$ energies 
for $^{196}$Os are predicted to lie below and beyond the $6^{+}_{1}$
level, respectively.  
For the exotic $^{192}$W and $^{194}$W nuclei, the present calculation
suggests a quite similar level pattern to their respective isotones, 
$^{194}$Os and $^{196}$Os.

\section{Summary}
 
To summarize, we have presented the predicted
excitation spectra and $B$(E2) ratios of exotic Os and W
isotopes with $N=114\sim 120$. 
Spectroscopic calculations have been carried out in terms of the IBM
Hamiltonian constructed by the constrained HFB calculations with Gogny-D1S EDF. 
We have examined the prolate-to-oblate shape/phase transition as 
functions of neutron number $N$ in the considered isotopic chains. 
The experimental trends of not only $g.s.$-band energies but also the
quasi-$\gamma$-bandhead $2^{+}_{\gamma}$ energy for Os isotopes is
reproduced well, suggesting that the $N=116$ nucleus is the softest in
$\gamma$. 
A similar pattern is predicted in W isotopes, while the evolution of
levels appears to occur more rapidly in W than in Os. 
Interestingly enough, all these results reflect to a good extent the
results of the underlying microscopic Gogny-HFB calculations.  
Lastly, let us comment on the form of the boson Hamiltonian in
Eq.~(\ref{eq:bh}). 
While this form may be rather simple, it determines the basic topology of the
PES, and is supposed to be the most 
relevant for the description of the low-lying structure at the present stage. 
On the other hand, the IBM-2 phenomenology considers additional interaction
terms as compared to those in Eq.~(\ref{eq:bh}). 
Some of these terms have a minor effect, but others might affect the spectroscopic
results in quantitative details as suggested in the structure of 
quasi-$\gamma$ bands in Fig.~\ref{fig:levelscheme}. 
It should be then very interesting to study in the future which parts of a 
more general boson Hamiltonian are crucial, as well as how they affect
the spectroscopic properties quantitatively.

\section*{Acknowledgments \label{sec:acknowledge}}
\addcontentsline{toc}{chapter}{Acknowledgments}

This work was support in part by grants-in-aid for Scientific Research
(A) 20244022 and No.~217368, and by the JSPS Core-to-Core program EFES. 
Author K.N. is support by JSPS Fellowship program. PHR, ZP and PDS acknowledge 
financial support from STFC(UK). The work of authors L.M.R, P.S  and R.R.
has been supported by MICINN (Spain) under
research grants 
FIS2008--01301, FPA2009-08958, and FIS2009-07277, as well as by 
Consolider-Ingenio 2010 Programs CPAN CSD2007-00042 and MULTIDARK 
CSD2009-00064. R.R. thanks Profs. P. M.Walker, J. \"Aysto and R.Julin
for encouraging discussions.


\begin{thebibliography}{99}
\addcontentsline{toc}{chapter}{Bibliography}

\bibitem{BM}
A.~Bohr and B.~R.~Mottelson, {\em Nuclear Structure}, (Benjamin, New York, 1969
	and 1975), Vols. I and II.  

\bibitem{RS}
P.~Ring and P.~Schuck, {\em The Nuclear Many-Body Problem}, (Springer, 
Berlin, 1980). 

\bibitem{Casten_nature}
P. Cejnar, J. Jolie, and R. F. Casten, Rev. Mod. Phys. \textbf{82}, 2155
	(2010).

\bibitem{Bender_review}
M.~Bender, P.-H.~Heenen, and P.-G.~Reinhard, Rev. Mod. Phys. \textbf{75}, 
	 121 (2003). 

\bibitem{Sk} 
T.~H.~R.~Skyrme, Nucl. Phys. \textbf{9}, 615 (1959). 

\bibitem{VB}
D.~Vautherin and D.~M.~Brink, Phys. Rev. C \textbf{5}, 626 (1972).

\bibitem{Go}
J.~Decharge, M.~Girod, and D.~Gogny, Phys. Lett. B \textbf{55}, 361 (1975). 

\bibitem{Vretenar-1} D.~Vretenar, A.~V.~Afanasjev, G.~A.~Lalazissis, 
and P.~Ring, Phys. Rep. \textbf{409}, 101 (2005).

\bibitem{CollSk}
P.~Bonche, J.~Dobaczewski, H.~Flocard, P.~-H.~Heenen, and J.~Meyer, 
	Nucl. Phys. \textbf{A510}, 466 (1990). 

\bibitem{CollGo}
J.-P.~Delaroche, M.~Girod, L.~Libert, H.~Goutte, S.~Hilaire, S.~Peru,
	 N.~Pillet, and G.~F.~Bertsch, Phys. Rev. C \textbf{81},  
	 014303 (2010). 

\bibitem{CollRHB}
Z.~P.~Li, T.~Nik$\check{\rm s}$i\'c, D.~Vretenar, and J.~Meng, Phys. Rev. C
	 \textbf{81}, 034316 (2010);
	 Z.~P.~Li, T.~Nik$\check{\rm s}$i\'c, D.~Vretenar, J.~Meng,
	G.~A.~Lalazissis, and 
	 P.~Ring, Phys. Rev. C \textbf{79}, 054301 (2009).  

\bibitem{AI}
A.~Arima and F.~Iachello, Phys. Rev. Lett. \textbf{35}, 1069 (1975); 
	  F.~Iachello and A.~Arima, {\em The interacting boson
	 model}, (Cambridge University Press, Cambridge, 1987). 

\bibitem{OAI}
T.~Otsuka, A.~Arima, F.~Iachello, and I.~Talmi, 
Phys. Lett. B \textbf{76}, 139 (1978); 
T.~Otsuka, A.~Arima, and F.~Iachello, Nucl. Phys. \textbf{A309}, 1 (1978). 

\bibitem{OCas}
T.~Otsuka, {\em Algebraic Approaches to Nuclear Structure}, 
(Harwood, Chur, 1993), ed. by R.F.~Casten, p. 195.

\bibitem{MO}
T.~Mizusaki and T.~Otsuka, Prog. Theor. Phys., Suppl. \textbf{125}, 97 (1997). 

\bibitem{nso}
K.~Nomura, N.~Shimizu, and T.~Otsuka, Phys Rev. Lett. \textbf{101}, 142501 (2008). 

\bibitem{nsofull}
K.~Nomura, N.~Shimizu, and T.~Otsuka, Phys Rev. C \textbf{81}, 
	044307 (2010). 

\bibitem{Shi06}  T.~Shizuma {\em et al.}, Eur.~Phys.~J. A \textbf{30}, 391 (2006). 

\bibitem{Pod00} Zs.~Podoly\'ak {\em et al.},  Phys. Lett. B \textbf{491}, 
	225 (2000).

\bibitem{Jolie2003HfHg}
J.~Jolie and A.~Linnemann, Phys. Rev. C \textbf{68}, 031301(R)
	(2003). 

\bibitem{Alk09} N.~Alkhomashi {\em et al.,} Phys. Rev. C \textbf{80}, 064308 (2009)
\bibitem{Reg08} P.H.~Regan {\em et al.,} Int.J.Mod.Phys. E17, Supplement
	1, 8 (2008).

\bibitem{Lane} G.J.~Lane {\em et al.,} Phys. Rev. C \textbf{82}, 051304
	(2010).  

\bibitem{Wheldon} C.~Wheldon {\em et al.}, 
        Phys. Rev. C \textbf{63}, 011304 (2001). 

\bibitem{Bond83}  P.D.~Bond, R.F.~Casten, D.D.~Warner, and D.~Horn,
	Phys. Lett. B 
                  \textbf{130}, 167 (1983).

\bibitem{stevenson} P.D.~Stevenson, M.P.~Brine, Zs.~Podoly\'ak, P.H.~Regan, 
        P.M.~Walker, and J.~Rikovska~Stone, Phys.Rev. C \textbf{72},
        047303 (2005). 

\bibitem{RaynerPt}
R.~Rodr\'iguez-Guzm\'an, P.~Sarriguren, L.~M.~Robledo, and 
	 J.~E.~Garc\'ia-Ramos, Phys. Rev. C \textbf{81}, 024310 (2010).  

\bibitem{gradient-2} L.~M.~Robledo, R.~Rodr\'{\i}guez-Guzm\'an, and 
	P.~Sarriguren, J. Phys. G: Nucl. Part. Phys. 36, 115104 (2009). 

\bibitem{DB_W}
P.~D.~Duval, B.~R.~Barrett, Phys. Rev. C \textbf{23}, 492 (1981).

\bibitem{nor}
K.~Nomura, T.~Otsuka, R.~Rodr\'iguez-Guzm\'an, L.~M.~Robledo, and 
	 P.~Sarriguren, Phys. Rev C \textbf{83}, 014309 (2011). 

\bibitem{D1S}
J.~F.~Berger, M.~Girod, and D.~Gogny, Nucl. Phys. \textbf{A428}, 23c (1984). 

\bibitem{coherent}
A.~E.~L.~Dieperink, O.~Scholten, and I.~Iachello,
	Phys. Rev. Lett. \textbf{44}, 1747 (1980). 

\bibitem{wavelet}
G.~Kaiser, {\em A Friendly Guide to Wavelets} (Boston, Birkh\"{a}ser, 1994).

\bibitem{SU3}
A.~Arima and F.~Iachello, Ann. Phys. (N.Y.) \textbf{111}, 201 (1978).

\bibitem{ibmmass}
K.~Nomura, T.~Otsuka, N.~Shimizu, and L.~Guo, arXiv:1011.1056. 

\bibitem{npbos}
T.~Otsuka and N.~Yoshida, JAERI-M report 85 (Japan Atomic
	 Energy Research Institute, 1985).

\bibitem{Pod09} Zs.~Podoly\'ak {\em et al.,} Phys. Rev. C \textbf{79}, 031305 (2009). 


\bibitem{data}
Brookhaven National Nuclear Data Center (NNDC)\\
http://www.nndc.bnl.gov/.

\bibitem{BE2}
C.~M.~Baglin, Nuclear Data Sheets \textbf{111}, 275 (2010); \\
C.~M.~Baglin, Nuclear Data Sheets \textbf{99}, 1 (2003); \\
B.~Singh, Nuclear Data Sheets \textbf{95}, 387 (2002); \\
B.~Singh, Nuclear Data Sheets \textbf{99}, 275 (2003); \\
C.~M.~Baglin, Nuclear Data Sheets \textbf{84}, 717 (1998); \\
B.~Singh, Nuclear Data Sheets \textbf{107}, 1531 (2006); \\
H.~Xiaolong, Nuclear Data Sheets \textbf{108}, 1093 (2007); \\
H.~Xiaolong, Nuclear Data Sheets \textbf{110}, 2533 (2007). 

 \bibitem{cubic}
K.~Heyde, P.~Van Isacker, M.~Waroquier, and J.~Moreau,
	 Phys. Rev. C \textbf{29}, 1420 (1984). 



\end{thebibliography}
\end{document}